\documentclass[12pt]{article}
\usepackage{cite}
\usepackage[pdftex]{hyperref}
\usepackage{amsmath, amsthm, amssymb, graphicx,slashed}
\usepackage{url}

\numberwithin{equation}{section}

\def\be{\begin{equation}}
\def\ee{\end{equation}}
\def\ba{\begin{align}}
\def\ea{\end{align}}
\def\beq{\begin{eqnarray}}
\def\eeq{\end{eqnarray}}
\def\a{\alpha}
\def\b{\beta}

\def\sgn{\text{sgn}}
\def\Im{\text{Im}}

\begin{document}

\begin{titlepage}

\vskip 1.5in
\begin{center}
{\bf\Large{Higher Poles and Crossing Phenomena\\ from Twisted Genera} }\vskip 0.5cm
{ Sujay K. Ashok${}^a$, Eleonora Dell'Aquila${}^{a}$ and Jan Troost${}^b$} \vskip
0.3in

 \emph{${}^{a}$}
\emph{ Institute of Mathematical Sciences \\
   C.I.T Campus, Taramani\\
   Chennai, India 600113\\ 
\vspace{.5cm}}

  \emph{\\${}^{b}$ Laboratoire de Physique Th\'eorique}\footnote{Unit\'e Mixte du CNRS et
     de l'Ecole Normale Sup\'erieure associ\'ee \`a l'universit\'e Pierre et
     Marie Curie 6, UMR
     8549.
}  \\
 \emph{ Ecole Normale Sup\'erieure  \\
 24 rue Lhomond \\ F--75231 Paris Cedex 05, France}
\end{center}
 \vskip 0.5in

\baselineskip 16pt

\begin{abstract}

 \vspace{.2in} We demonstrate that Appell-Lerch sums with higher
  order poles as well as their modular covariant completions arise as
  partition functions in the cigar conformal field theory with
  worldsheet supersymmetry. The modular covariant derivatives of the
  elliptic genus of the cigar give rise to operator insertions
  corresponding to (powers of) right-moving momentum, left-moving
  fermion number, as well as a term corresponding to an ordinary zero
  mode partition sum. To show this, we demonstrate how the
  right-moving supersymmetric quantum mechanics (and in particular the
  Hamiltonian and spectral density) depend on the imaginary part of
  the chemical potential for angular momentum. A consequence of our
  analysis is that varying the imaginary part of the chemical
  potential for angular momentum on the cigar gives rise to a
  wall-crossing phenomenon in the bound state contribution to the
elliptic genus, while the full elliptic genus is a continuous
function of the chemical potential.
\end{abstract}
\end{titlepage}
\vfill\eject

\tableofcontents

\section{Introduction}
Two-dimensional conformal field theories are of physical as well as
mathematical interest.  One of many ways in which these physical
theories connect to mathematics is through the calculation of their
elliptic genera.  In the case
of compact target space manifolds for two-dimensional non-linear
sigma-models, the elliptic genera
 capture a plethora of Dirac indices on
symmetrized and anti-symmetrized tangent vector bundles of the target space
\cite{Witten:1986bf}.

Elliptic genera can be defined for two-dimensional conformal field
theories which have at least one right-moving supercharge. For the
right-movers, one mimics the definition of the Witten index, while
computing the partition sum for left-movers, twisted by all charges which
commute with the right-moving supercharge.

In this paper, we discuss further aspects of the elliptic genus in a
two-dimensional supersymmetric cigar conformal field theory, with an
$N=(2,2)$ superconformal symmetry algebra. The elliptic genus, twisted by a global angular momentum charge $P$, is defined as the
following trace over the Ramond-Ramond sector Hilbert space ${\cal H}$:
\be\label{EGdef}
\chi(\tau, \a, \b) = \text{Tr}_{\cal H}(-1)^{F_L +F_R} q^{L_0-\frac{c}{24}}z^{J^R_0}y^{P}\,,
\ee
where we used the notation $q=e^{2\pi i \tau}$, $z=e^{2\pi i \a}$ and
$y=e^{2\pi i \b}$.  The operator $J_0^R$ measures the left moving $U(1)$
$R$ charge while $P$ measures the global angular momentum of the
states\footnote{Following thermodynamics nomenclature, we will often refer to $\a$ and $\b$ as chemical
  potentials.}. 
  
The elliptic genera of the supersymmetric Liouville and cigar theories
were calculated in \cite{Troost:2010ud, Eguchi:2010cb, Ashok:2011cy}
using the path integral formulation of these theories. In particular,
we will look at the three variable Jacobi form analyzed in
\cite{Ashok:2011cy} and generalize this result by deriving a path
integral expression for the elliptic genus with complex chemical
potentials for the R-charge and the global charge.  This
generalization is natural from a mathematical point of view given that
elliptic genera are Jacobi forms, and their arguments are transformed
within the set of complex numbers under elliptic and modular transformations.

Once the chemical potentials are complexified, it becomes straightforward to take modular covariant
derivatives. We will show that this allows us to find
physical models for certain Appell-Lerch sums with higher order poles
\cite{DMZ}, within the cigar or Liouville conformal field theory. The
covariant derivatives correspond to operator insertions of
right-moving momentum and left-moving fermion number, as well as 
a contribution from a zero mode partition sum.  The right-moving momentum is
not strictly conserved, but is a good asymptotic quantum number that
can be used to label states in partition sums. We will show that a
modification of the supersymmetric quantum mechanics for right-moving primaries is
induced by the insertions, which changes the difference in spectral
densities arising in the integral over the continuum.

As a result of our analysis, we will encounter a wall-crossing
phenomenon in a simple two-dimensional superconformal field theory.
By varying the imaginary part of the angular momentum chemical
potential $\b$, some states are subtracted from the discrete part of the
spectrum, as coded in the holomorphic part of the elliptic genus, while other
states are added to the bound state spectrum. The continuum contribution 
exhibits a complementary behaviour such that the  full elliptic genus is
continuous.

Another motivation for our analysis comes from the widening range of applicability of mock modular forms
in physics. Most of the early applications focused on restoring a duality or
modular invariance \cite{Zwegers, Zagier} in identifiable holomorphic contributions to e.g.
superconformal characters \cite{Eguchi:1987wf,Semikhatov:2003uc,Eguchi:2008gc} or supergravity
partition functions affected by wall-crossing \cite{Manschot:2009ia}. The modular completion
was shown to arise naturally in the context of
 the superconformal coset partition function calculation
\cite{Troost:2010ud}, which allowed for many generalizations in the two-dimensional
realm, and applications to the physics of two-dimensional and higher-dimensional
black holes in string theory
\cite{Giveon:2013hsa,Giveon:2014hfa}. Space-time indices inherit mock modularity
of the worldsheet indices \cite{Harvey:2013mda}. Still, most of the mock modularity of
space-time indices (see e.g \cite{DMZ,Cardoso:2013ysa})
remains poorly understood, and providing microscopic models for 
mock modular forms in two dimensions (e.g. on the worldsheet of an (effective) string) 
may well be instrumental in identifying the relevant physics in space-time.

Our paper is structured as follows. We start in section
\ref{pathintegral} by proposing the path integral expression for the
elliptic genus with chemical potentials taking values in the complex
plane.  We then perform the traditional transform to the Hamiltonian
form in order to recover known results in the mathematics literature, and interpret the result in terms
of a modified right-moving supersymmetric quantum mechanics. We then
exploit the result 
to exhibit a wall
crossing phenomenon as a function of the imaginary part of the
chemical potential for angular momentum.\footnote{We will
use the word wall crossing to refer to a jump in the bound state
spectrum arising when a modulus crosses a particular value.} In section \ref{higherorder}
we further apply the result to consider modular covariant derivatives
of these results.  These will serve to model higher order Appell-Lerch
sums and to give a direct conformal field theory interpretation of
their modular completions, in particular in a Hamiltonian form. 
We conclude in section \ref{conclusions} with a discussion and suggestions
for generalizations and applications.

\section{Path integral elliptic genera}
\label{pathintegral}

Elliptic genera, defined as twisted partition functions on tori, have natural
elliptic and modular properties; they are Jacobi forms \cite{Kawai:1993jk}. The 
modular transformation
properties of the three variable elliptic genus of the $N=2$ Liouville conformal
field theory we will study are:
\begin{align}
\label{transfoprop}
\chi(\tau+1, \a, \b) &= \chi(\tau, \a, \b) \cr
\chi(-\frac{1}{\tau}, \frac{\a}{\tau},\frac{\b}{\tau}) &= e^{\frac{c}{3}\frac{\pi i \a^2}{\tau}-\frac{2\pi i \a\b}{\tau}} \chi(\tau, \a, \b)\, .
\end{align}
where $c=3+6/k$ is the central charge of the theory and we will take the level $k$ to be a positive integer.
The genus moreover has periodicity properties in $\alpha$ under shifts
by integer multiples of $k(1,\tau)$ and in $\beta$ under shifts by multiples
of $1$ and $\tau$.
Since these elliptic transformations and the modular transformations in (\ref{transfoprop}) shift and rescale chemical
potentials by the complex parameter $\tau$, 
it is natural to study elliptic genera for complex chemical
potentials.  This demands a slight generalization of the analysis
in \cite{Troost:2010ud, Eguchi:2010cb, Ashok:2011cy}. 
Further physical motivations
for this generalization will become clear in the course of the paper.

\subsection{The path integral expression}
In an earlier work \cite{Ashok:2011cy}, we obtained the following expression for the elliptic genus of the cigar superconformal field theory:
\begin{align}
\chi_{cos}(\tau, \a, \b) 
= \int_0^1 d s_{1,2}\  
\frac{\theta_{11} (\tau,s_1\tau +s_2 - \frac{k+1}{k} \alpha+\beta )}{
\theta_{11}(\tau, s_1 \tau+ s_2 - \frac{\alpha}{k} + \beta )} 
\sum_{m,w \in \mathbb{Z}}
e^{- 2 \pi i s_2 w + 2 \pi i s_1 (m-\alpha)}\ 
e^{ - \frac{\pi}{k \tau_2} | m-\alpha+ w \tau |^2} \, . \label{1101result}
\end{align}
Modularity and ellipticity of the elliptic genus were exhibited in  \cite{Ashok:2011cy}
 using this expression. This is therefore the expression that generalizes naturally to
the case of complex chemical potentials $\alpha=\alpha_1+\tau \alpha_2$ and
$\beta=\beta_1 + \tau \beta_2$.
A double Poisson resummation (and combining holonomy variables and winding numbers)
 produces the path integral form of the elliptic genus:
\begin{eqnarray}
\chi_{cos}(\tau, \a, \b) =
k\, \int_{-
\infty}^{+
\infty} ds_{1} ds_{2}\ 
\frac{\theta_{11} (\tau,s_1\tau +s_2 - \frac{k+1}{k} \alpha+\beta )}{
\theta_{11}(\tau, s_1 \tau+ s_2 - \frac{\alpha}{k} + \beta )}\ 
e^{ - \frac{k \pi}{ \tau_2} | s_1  \tau + s_2  |^2} \ 
 e^{-2\pi i\a_2(s_1 \tau + s_2)} \, .
\label{pathintegralformula}
\end{eqnarray}
Compared with \cite{Ashok:2011cy}, here we have an additional factor
whose exponent is proportional to $\alpha_2\,$, since we have now
allowed for complex $\alpha$. A path integral derivation of the formula is
given in Appendix \ref{cosetpathintegral}. It is interesting to
observe that this extra factor can be combined in the following way by
shifting the integration variable by $\b-\frac{\a}{k}\,$:
\begin{eqnarray}
\chi_{cos}(\tau, \a, \b) =
 k\, \int_{-
\infty}^{+
\infty} ds_{1} ds_{2}
\frac{\theta_{11} (\tau,s_1\tau +s_2 -  \alpha)}{
\theta_{11}(\tau, s_1 \tau+ s_2)} 
e^{ - \frac{k \pi}{ \tau_2} ( s_1  \tau + s_2 +\frac{\a}{k}-\b )(s_1\bar{\tau} +s_2 +\frac{\a}{k} -\bar{\b}) } 
\end{eqnarray}
This manipulation shows explicitly that the elliptic genus is
independent of $\bar{\a}$ and depends only holomorphically on the $\a$
variable\footnote{We would like to thank Sameer Murthy for helpful
  discussions on this point.}. Interestingly the elliptic genus is not
holomorphic in $\beta$. One either sees this by studying the exponent, or by noting
that there is a pole in $\beta$. This property will prove crucial in
exhibiting a wall crossing phenomenon in the conformal field theory.

An equivalent, equally useful way to write the path integral
answer, is in terms of holonomies $s_{1,2}$ in the interval $[0,1]$ and
integer winding numbers $m$ and $w$:
\begin{multline}\label{chicoss1s2}
\chi_{cos}(\tau, \a, \b) =
k\, \sum_{n,m} \int_0^1 ds_{1} ds_{2}\ 
\frac{\theta_{11} (\tau,s_1\tau +s_2 - \alpha )}{
\theta_{11}(\tau, s_1 \tau+ s_2 )}\ e^{ 2 \pi i \alpha n}  \cr \times e^{ - \frac{k \pi}{ \tau_2} | (n+ s_1 ) \tau + m+ s_2 +\frac{\alpha}{k} -  \beta |^2}
e^{-2\pi i\a_2(m +s_2+\frac{\alpha}{k} + \tau(n + s_1)- \b)} \, .
\end{multline}
We have shifted the $\alpha$ and $\beta$ dependence of the bosons
into the exponential factor. This form of the elliptic genus path
integral arises naturally from the derivation in terms of a gauged
linear sigma model \cite{Murthy:2013mya,Ashok:2013pya}. Again, we have
generalized the result to complexified $\alpha$ and $\b$.

\subsection{Modularity and periodicity}

Another strong argument for the generalized path integral expression in 
\eqref{pathintegralformula} is that it can now be checked directly, using equation
(\ref{pathintegralformula}), to be modular and elliptic.
Indeed, invariance under the modular $T$ transformation is easily
established by shifting $s_2$ appropriately.  Under the
modular $S$-transformation, the variables transform as
\begin{align}
\tau\rightarrow -\frac{1}{\tau} \qquad \a \rightarrow \frac{\a}{\tau} \qquad  
\b\rightarrow \frac{\b}{\tau} \qquad s_2\rightarrow s_1 \qquad s_2\rightarrow -s_1 
\qquad \a_2\longrightarrow \tau\, \a_2 - \a \,.
\end{align}
Using these and the appropriate modular properties of the theta function (see Appendix \ref{usefulformulae}), 
one can check that the integrand picks up an exponential factor 
that matches the modular transformation in
equation \eqref{transfoprop}, with the central charge $c=3+6/k$
of the supersymmetric cigar theory.

The periodicity properties as well can be checked
directly on the path integral expression \eqref{pathintegralformula}. 
Consider for concreteness
the shift 
$ \a \rightarrow \a + k\tau \,.$
The absolute value term in the exponent of \eqref{pathintegralformula} is unchanged by this transformation; the last ($\a_2$ dependent) factor picks up a contribution
$ e^{-2\pi i k (s_1\tau + s_2)}\,$.
Combining this with the elliptic property of the theta function recorded in equation
\eqref{thetamodell}, the integrand of the elliptic genus picks up a
combined factor:
\be
\frac{(-1)^{k+1}q^{-\frac{(k+1)^2}{2}} q^{(k+1)s_1}e^{2\pi i s_2 (k+1)} z^{-\frac{(k+1)^2}{k}} y^{k+1}}{(-1)q^{-\frac{1}{2}} q^{s_1}e^{2\pi i s_2 } z^{-\frac{1}{k}} y}\, e^{-2\pi i k (s_1\tau + s_2)} 
= (-1)^k q^{-\frac{k^2+2k}{2}}z^{-(k+2)} y^k \, .
\ee
Using the value of the central charge, we can write the
elliptic property of the elliptic genus as
\be
\chi_{cos}(\tau, \a +k\tau, \b) = (-1)^{\frac{c}{3}k} e^{-\frac{\pi i c}{3} (k^2 \tau + 2k \a)} e^{2\pi i \b k}\chi_{cos}(\tau, \a, \b) \,.
\ee
Along these lines we can also check the ellipticity under shift of
$\alpha$ by integer multiples of $k$, and shifts of $\b$ by integer
multiplies of $1$ and $\tau$. The resulting properties are those
recorded in \cite{Ashok:2011cy}. Here, we derived these properties
directly from the path integral expression which is now
valid for complexified chemical potentials $\alpha$ and $\beta$.

\subsection{The Liouville elliptic genus}
The cigar elliptic genus is related to that of $N=2$ Liouville theory
by a $\mathbb{Z}_k$ orbifold \cite{
  Eguchi:2010cb,Ashok:2011cy}. Thus, we find for the Liouville
elliptic genus:
\begin{multline}\label{zwegersresum}
\chi_L(\tau, \a, \b)
=\sum_{n,m} \int_0^1 ds_{1} ds_{2}
\frac{\theta_{11} (\tau,s_1\tau +s_2 - \alpha )}{
\theta_{11}(\tau, s_1 \tau+ s_2 )} e^{ 2 \pi i \alpha \frac{n}{k}}
e^{ - \frac{k \pi}{ \tau_2} | (\frac{n}{k}+ s_1 ) \tau + (\frac{m}{k}+ s_2 ) + \frac{\alpha}{k} -  \beta |^2} 
\cr
\times e^{-2\pi i\a_2((\frac{n}{k}+ s_1 ) \tau + (\frac{m}{k}+ s_2 ) +\frac{\alpha}{k} -  \beta)}\, .
\end{multline}
The Liouville elliptic genus is most simply related
\cite{Troost:2010ud} to the single pole Appell-Lerch sum studied in 
the mathematics literature \cite{Zwegers}. Since we wish to compare our results with this
literature, we work in Liouville theory in what follows. 
All our statements apply, mutatis mutandis, to orbifolds of Liouville theory,
tensor product theories, et cetera. In particular, they have straightforward
equivalents for the two-dimensional black hole superconformal field theory.

\subsection{The Hamiltonian perspective}

It is instructive to understand the sum over states coded in the path integral result. In order
to reach this perspective, one needs to switch from a Lagrangian to a Hamiltonian viewpoint.
This can be achieved after a number of technical steps that were performed in
\cite{Troost:2010ud, Eguchi:2010cb, Ashok:2011cy}; the generalization of these steps
to complexified chemical potentials $\alpha$ and $\beta$ is  detailed in appendix
\ref{lagrangiantohamiltonian}. The final result is a sum of two terms, written out in \eqref{appchiholo} and \eqref{remainderintegral}; the first one is a holomorphic, right-moving ground state contribution to the elliptic genus of the form:
\begin{multline}
\chi_{L,hol}(\tau, \a, \b) =
\frac{ i \theta_{11}(\tau,\alpha)}{\pi\eta^3}
\sum_{m,v,w}\left[\int_{\mathbb{R}}-\int_{\mathbb{R}-\frac{ik}{2}} \right]
\frac{ds}{ 2 i s + v - k \beta_2}
S_{m+kw-1}
\cr
\times \ z^{v/k-2w} y^{kw} q^{-vw+kw^2}  (q \bar{q})^{(is + \frac{v}{2} - \frac{k \beta_2}{2})+\frac{s^2}{k}+\frac{ (v-k \beta_2)^2}{4k}} \, , 
\label{differencecontours}
\end{multline}
where $v=n+kw$ represents the right-moving momentum on the asymptotic
circle (in terms of the angular momentum $n$ and winding number
$w$) and $S_r$ is a sum introduced in appendix \ref{usefulformulae}. The contour integral picks up poles corresponding to right-moving
ground state contributions, which are holomorphic.
 The new ingredient, compared to the earlier works, is the
$\beta_2$ dependence. We observe that if $k \beta_2$ is not an
integer, the contour integrals are unambiguously defined. If $k\b_2$ is an
integer, we define our integrals parallel to the real line to be
shifted slightly, by adding a small imaginary part $\epsilon$
to the integration contour.  In all cases, the
contours are taken such that we sum the right-moving momentum $v$ over a range of integers, the
highest of which is the integer $[k \beta_2]$ smaller or equal to $k \beta_2$, 
and the lowest of which is that integer minus
$k-1$. In total, there are $k$ integer valued right-moving momenta $v$ between the two contours.
We obtain the expression:
\be
\chi_{L,hol}(\tau, \a, \b)= -\sum_{m,w} \sum_{v=-(k-1)+[k \beta_2]}^{[k \beta_2]}(-1)^m q^{(m-1/2)^2/2} z ^{m-1/2}
S_{m+kw-1}
z^{v/k-2w} y^{kw} q^{kw^2-vw} \, .
\label{pickingpoles}
\ee
If we define $m=-\tilde{m}+1$ and use the identity \eqref{thetabypole} we find that the holomorphic
part of the elliptic genus is given by:
\begin{align}
\chi_{L,hol}(\tau, \a, \b)&= \frac{i\theta_{11}(\tau, -\a)}{\eta^3(\tau)}\sum_{w} \frac{z^{-2w} y^{kw} q^{kw^2}}{1-z^{-1}q^{kw}}   
( z^{\frac{1}{k}} q^{-w})^{[k \beta_2]}
\sum_{v=0}^{k-1}(z^{-1}q^{w})^{v}\cr
&=  \frac{i\theta_{11}(\tau, -\a)}{\eta^3(\tau)}\sum_{w} \frac{z^{-2w} y^{kw} q^{kw^2}}{1-z^{-\frac{1}{k}}q^{w}}  ( z^{\frac{1}{k}} q^{-w})^{[k \beta_2]} \cr
&=  z^{[k \beta_2]/k} \frac{i\theta_{11}(\tau, -\a)}{\eta^3(\tau)}\sum_{w} \frac{ ( z^{-2} y^{k} q^{- [k\beta_2]})^w 
q^{kw^2}}{1-z^{-\frac{1}{k}}q^{w}}  \,.
\label{holomorphicpiece}
\end{align}
This is a sum over extended $N=2$ superconformal algebra characters based on Ramond ground
states of R-charge $ 1/2-l/k + [k \beta_2]/k$ where $l$ takes values in the set
$l=0,1,\dots,k-1$. This can be seen from the first line, where we identify $w$ as the spectral
flow summation variable.

There is also a non-holomorphic contribution to the elliptic genus, which arises 
from the difference in spectral densities for fermionic and bosonic
right-moving primaries:
\be
\chi_{L,rem}(\tau, \a, \b) = 
\frac{i\theta_{11}(\tau,-\a)}{\pi\eta^3(\tau)}\sum_{v,w}\int_{\mathbb{R}} \frac{ds}{ 2 i s + v - k \beta_2}
z^{v/k-2w} y^{kw} q^{kw^2-vw}  (q \bar{q})^{\frac{s^2}{k}+\frac{ (v-k \beta_2)^2}{4k}} \, .
\label{remainderpiece}
\ee
These two terms, in \eqref{holomorphicpiece} and \eqref{remainderpiece} sum to the modular completion of the Appell-Lerch sum
analyzed in the mathematics literature
\cite{Zwegers}, as we show in the next section. 

\subsection{The relation to completed Appell-Lerch sums}
\label{ALmath}
In this subsection, we wish to show how the path integral result for the
Liouville elliptic genus for complexified arguments, rewritten in the
Hamiltonian form, relates
to the mathematics of modularly completed Appell-Lerch
sums \cite{Zwegers}.
\subsubsection{Review of the completed Appell-Lerch sum $\widehat{A}$}
The
holomorphic Appell-Lerch sum is defined
as\footnote{In \cite{Zwegers} the Appell-Lerch sum is denoted
  $A_{2k}$. Since we go on to define a series of higher pole
  Appell-Lerch sums, we use a notation close (but not identical) to \cite{DMZ}.}
\be\label{ALsum1}
{\cal A}_{1,k}(\tau, u, v) = a^{k}\sum_{n\in \mathbb{Z}} \frac{q^{kn(n+1)}b^n}{1-aq^n}
\, .
\ee
In the conventions of \cite{Zwegers}, the three
variables are denoted by $q=e^{2 \pi i \tau}$, $a=e^{2\pi i u}$ and $b=e^{2\pi i v}$. 
The remainder of this Appell-Lerch sum, which, 
when added to the holomorphic
part \eqref{ALsum1} leads to a Jacobi form, is given by \cite{Zwegers}
\be
{\cal R}_{1, k}(\tau, u,v)=\frac{i}{4k}a^{\frac{2k-1}{2}}\sum_{m(mod)2k}\theta_{11}\left(\frac{v+m}{2k}+\frac{(2k-1)\tau}{4k};\frac{\tau}{2k}\right) R\left(u-\frac{v+m}{2k}-\frac{(\tau(2k-1)}{4k};\frac{\tau}{2k}\right)
\ee
where the function $R$ is defined as:
\be\label{Rdef}
R(u;\tau) = \sum_{\nu\in\mathbb{Z}+\frac{1}{2}}
\left(\sgn(\nu)-\text{Erf}\left[\sqrt{2\pi \tau_2}\left(\nu+\frac{\Im(u)}{\tau_2}\right)\right] \right)
(-1)^{\nu-\frac{1}{2}}a^{-\nu}q^{-\frac{\nu^2}{2}} \, .
\ee
The sum 
\be
\widehat{\cal A}_{1,k}(\tau, u,v) = {\cal A}_{1, k}(\tau, u,v) + {\cal R}_{1, k}(\tau, u,v) 
\ee
satisfies good modular and elliptic properties \cite{Zwegers}:
\begin{align}\label{ALmodularelliptic}
{\widehat {\cal A}}_{1, k}(\tau, u+1, v) &= {\widehat {\cal A}}_{1, k}(\tau, u,v) \qquad \qquad\qquad
{\widehat {\cal A}}_{1, k}(\tau, u, v+1) = {\widehat {\cal A}}_{1, k}(\tau, u,v)  \\
{\widehat {\cal A}}_{1,k}(\tau, u+\tau, v) &= a^{2k}b^{-2k}q^{k}{\widehat {\cal A}}_{1, k}(\tau, u,v) \qquad 
{\widehat {\cal A}}_{1, k}(\tau, u, v+\tau) = a^{-1}{\widehat {\cal A}}_{1, k}(\tau, u,v) \nonumber \\
{\widehat {\cal A}}_{1, k}(\tau+1, u, v) &= {\widehat {\cal A}}_{1, k}(\tau, u,v) \qquad \qquad\qquad
{\widehat {\cal A}}_{1, k}(-\frac{1}{\tau},\frac{u}{\tau}, \frac{v}{\tau}) = \tau\, e^{\frac{2\pi i }{\tau}(vu-ku^2) }{\widehat {\cal A}}_{1, k}(\tau, u,v)\,. \nonumber 
\end{align}
The map to the 
conformal field theory variables used earlier is:
\be\label{map2}
u= -\frac{\a}{k} \qquad v = -2\a - k\tau+k\b- \tau [k \beta_2] \,,
\ee
Through this map, we can rewrite the Appell-Lerch sum in the form:
\be
{\cal A}_{1, k}= z^{-1} 
\sum_{w\in \mathbb{Z}} \frac{q^{kw^2} (z^{-2} y^{k} q^{- [k \beta_2]})^w}{1-z^{-\frac{1}{k}}q^w}
\, ,
\ee
with the usual notation $q=e^{2\pi i \tau}$, $z=e^{2\pi i \a}$ and $y=e^{2\pi i \b}$. 
Note that the choice of map (2.19) gives as an immediate match between the holomorphic part of the Appell-Lerch sum and the holomorphic part of the elliptic genus. Matching the remainder terms is less straightforward and it will be the object of the rest of this section.

Applying the map \eqref{map2} to the remainder yields 
\be\label{R1kinter}
{\cal R}_{1, k}=\frac{i}{4k}z^{-1+\frac{1}{2k}}\sum_{m(mod)2k}
\theta_{11}\left(-\frac{\a}{k}+\frac{\b}{2}- \frac{\tau [k \beta_2]}{2k}+\frac{m}{2k}-\frac{\tau}{4k} ;\frac{\tau}{2k}\right) 
R\left(\frac{\tau}{4k}-\frac{\b}{2}+ \frac{\tau [k \beta_2]}{2k}-\frac{m}{2k};\frac{\tau}{2k}\right)\, .
\ee
With these arguments, the function $R$ in \eqref{Rdef} evaluates to
\begin{align}\label{Rexp2}
R
&= \sum_{\nu\in\mathbb{Z}+\frac{1}{2}}
(-1)^{\nu-\frac{1}{2}}y^{\frac{\nu}{2}} q^{-\nu [k \beta_2]/2k}
q^{-\frac{\nu^2}{4k} -\frac{\nu}{4k}}e^{\frac{2\pi i \nu m}{2k}}
\left(\sgn(\nu)-
\text{Erf}
\left[
\sqrt{\frac{\pi \tau_2}{k}}\left(\nu+\frac{1}{2}-\frac{\Im(k\, \b - \tau[k \beta_2])}{\tau_2}\right) 
\right]
\right)\cr
&=-y^{-\frac{1}{4}} q^{\frac{1}{16k}}\sum_{r\in \mathbb{Z}}
(-1)^{r}y^{\frac{r}{2}}
q^{-r [k \beta_2]/2k}
q^{[k \beta_2]/4k}
q^{-\frac{r^2}{4k}}e^{\frac{2\pi i (r-\frac{1}{2}) m}{2k}} \cr
&\hspace{4cm}
\times \left(\sgn(r-\frac{1}{2})-
\text{Erf}
\left[
\sqrt{\frac{\pi \tau_2}{k}}\left(r-\frac{ \Im(k\, \b - \tau[k \beta_2])}{\tau_2}\right) 
\right]
\right)\,. 
\end{align}
The $\theta_{11}$ function that appears in \eqref{R1kinter} evaluates to
\be\label{thetaexp2}
\theta_{11}\left(-\frac{\a}{k}+\frac{\b}{2}-\frac{\tau [k \beta_2]}{2k} 
+\frac{m}{2k}-\frac{\tau}{4k} ;\frac{\tau}{2k}\right)=
i
\sum_{n\in \mathbb{Z}} (-1)^n q^{\frac{n^2}{4k}}q^{-\frac{1}{16k}}
z^{\frac{n}{k}-\frac{1}{2k}}
y^{-\frac{n}{2}+\frac{1}{4}}
q^{ \frac{\tau [k \beta_2]}{2k} (n-\frac{1}{2})} 
e^{\frac{2\pi i  m}{2k}(\frac{1}{2}-n)} \, . 
\ee
Multiplying the two expressions in equations \eqref{Rexp2} and \eqref{thetaexp2}, all the constant exponents of $q$, $y$ and $z$ cancel except
for one factor of $z$. Furthermore, the variable $m$ can be summed to give
\be
\sum_{m\, (mod)\, 2k} e^{\frac{2\pi i m}{2k}(r-n)} = 2k\, \delta_{n-r+2k \mathbb{Z}}
\, .
\ee
Substituting the solution to the constraint equation $n = r+2kj$,
where $j\in \mathbb{Z}$, one notices that the summation over $j$
results in a theta function with indices $(r,k)$ (see equation
\eqref{thetark2}). Gathering all these results leads to a more compact
expression for the remainder term:
\begin{multline}
{\cal R}_{1,k} = \frac{z^{-1}}{2}\sum_{r\in \mathbb{Z}}\left(\sgn(r-\frac{1}{2})-
\text{Erf}
\left[
\sqrt{\frac{\pi \tau_2}{k}}\left(r-\frac{\Im(k\b- \tau[k \beta_2])}{\tau_2}\right) 
\right]
\right)\, y^{\frac{r}{2}}\, 
q^{- \frac{r [k \beta_2]}{2k}}\, 
q^{-\frac{r^2}{4k}}
\cr
\times\, \theta_{r,k}\left(\tau, \frac{\a}{k}-\frac{\b}{2}+\frac{\tau [k \beta_2]}{2k}\right)
\, .
\end{multline}
Flipping the sign of $r$ and using the fact that both the sign and error functions are odd, we obtain:
\begin{multline}\label{finalzwegers2}
{\cal R}_{1, k} = -\frac{z^{-1}}{2}\sum_{r\in \mathbb{Z}}\left(\sgn(r+\frac{1}{2})-
\text{Erf}
\left[
\sqrt{\frac{\pi \tau_2}{k}}\left(r+\frac{\Im(k\b- \tau[k \beta_2])}{\tau_2}\right) 
\right]
\right)
\cr
y^{-\frac{r}{2}}\, 
q^{ \frac{r [k \beta_2]}{2k}}\, 
q^{-\frac{r^2}{4k}}\,  \theta_{r,k}\left(\tau, -\frac{\a}{k}+\frac{\b}{2}-\frac{\tau [k \beta_2]}{2k}\right)\,. 
\end{multline}
Consider the argument of the error function: recalling that 
$\b = \b_1 + \tau \b_2$, we find that
\be
r+\frac{1}{\tau_2}\Im(k\beta - \tau [k \beta_2])
= r+ (k\beta_2-[k \beta_2]) 
\ee
We therefore obtain the part of $k \beta_2$ that lies in the interval 
$(0,1)$, which we will denote by $\gamma_2=k \beta_2 -[k \beta_2]$.
We then have:
\be\label{finalzwegers3}
{\cal R}_{1,k} = -\frac{z^{-1}}{2}\sum_{r\in \mathbb{Z}}\left(\sgn(r+\frac{1}{2})-
\text{Erf}
\left[
\sqrt{\frac{\pi \tau_2}{k}}(r+\gamma_2) 
\right]
\right)y^{-\frac{r}{2}}\,
q^{ \frac{r [k \beta_2]}{2k}}\,
q^{-\frac{r^2}{4k}}\, \theta_{r,k}\left(\tau, -\frac{\a}{k}+\frac{\b}{2}-\frac{[\tau k \b_2]}{2k}\right)
\ee
We now add and subtract $\sgn(r+\gamma_2)$ in the parenthesis and observe that
\be
\sum_{r\in \mathbb{Z}}(\sgn(r+\frac{1}{2}) -\sgn(r+\gamma_2))f(r) = 0\, ,
\ee
except when $\gamma_2=0$ (due to the convention $\sgn(0)=0$).
Restricting to $\gamma_2 \neq 0$ -- the special case can be treated analogously --, 
this leads to 
\be
{\cal R}_{1,k} = -\frac{z^{-1}}{2}\sum_{r\in \mathbb{Z}}
\left(\sgn(r+\gamma_2)-
\text{Erf}
\left[
\sqrt{\frac{\pi \tau_2}{k}}(r+\gamma_2) 
\right]
\right)y^{-\frac{r}{2}}\,
q^{ \frac{r [k \beta_2]}{2k}}\,
q^{-\frac{r^2}{4k}}\, 
\theta_{r,k}\left(\tau, -\frac{\a}{k}+\frac{\b}{2}-\tau\frac{[k \b_2]}{2k}\right)
\, .
\ee
Using the integral \eqref{signminuserf} we
can write the remainder ${\cal R}$ as 
\be
{\cal R}_{1, k} = -\frac{z^{-1}}{\pi}
\sum_{r\in \mathbb{Z}} \int_{\mathbb{R} + i\epsilon}
\frac{ds}{2is + r+\gamma_2} (q\bar{q})^{\frac{s^2}{k} +\frac{(r+\gamma_2)^2}{4k}}\,
y^{-\frac{r}{2}}\, 
q^{ \frac{r [k \beta_2]}{2k}}\,
q^{-\frac{r^2}{4k}}\, \theta_{r,k}\left(\tau, -\frac{\a}{k}+\frac{\b}{2}-\tau\frac{[k \b_2]}{2k}\right) \,.
\ee
We use the definition of the theta function in \eqref{thetark2}, setting $j=w$ 
and also relabel $r=-v$ to obtain:
\be
{\cal R}_{1, k} = -\frac{z^{-1}}{\pi}
\,\sum_{v\in \mathbb{Z}}\sum_{w\in \mathbb{Z}}
q^{kw^2-vw}
z^{-2w+\frac{v}{k}}
y^{kw}
q^{- w [k \beta_2]} 
\int_{\mathbb{R}}\frac{ds}{2is -v+\gamma_2} (q\bar{q})^{\frac{s^2}{k} +\frac{(v-\gamma_2)^2}{4k}}
\label{Zwint2}
\ \, .
\ee
Finally, we shift the variable $v$ to $v=\tilde{v}-[k \beta_2]$ and obtain (after dropping
the tilde):
\be
{\cal R}_{1,k} = \frac{z^{-1-\frac{[k \beta_2]}{k}}}{\pi}
\,\sum_{v\in \mathbb{Z}}\sum_{w\in \mathbb{Z}}
q^{kw^2-vw}
z^{-2w+\frac{v}{k}}
y^{kw}
\int_{\mathbb{R}}\frac{ds}{2is +v-k \beta_2} 
(q\bar{q})^{\frac{s^2}{k} +\frac{(v-k \beta_2)^2}{4k}}
\label{Zwint3}
\ \, .  \ee 
Here we have also flipped the sign of $s$ in the integral that picks up an extra sign. To summarize, the modularly completed Appell-Lerch sum ${\widehat {\cal A}}_{1, k}$ of \cite{Zwegers} can be written as the sum of the following holomorphic and remainder terms 
\begin{align}\label{ALintermediate}
{\cal A}_{1, k}
&= 
z^{-1} 
\sum_{w\in \mathbb{Z}} \frac{q^{kw^2} (z^{-2} y^{k} q^{- [k \beta_2]})^w}{1-z^{-\frac{1}{k}}q^w}
\cr
{\cal R}_{1,k}
&=  \frac{z^{-1-\frac{[k \beta_2]}{k}}}{\pi}
\,\sum_{v\in \mathbb{Z}}\sum_{w\in \mathbb{Z}}
q^{kw^2-vw}
z^{-2w+\frac{v}{k}}
y^{kw} 
\int_{\mathbb{R} }\frac{ds}{2is+v-k \beta_2} 
(q\bar{q})^{\frac{s^2}{k} +\frac{(v-k \beta_2)^2}{4k}} \,.
\end{align}
If we further multiply this result by the appropriate overall factor 
we obtain precisely the elliptic genus of the supersymmetric Liouville theory, as in equations (\ref{holomorphicpiece}) and (\ref{remainderpiece}).
Moreover, we can use the periodicity properties of the completed Appell-Lerch sum ${\widehat {\cal A}}_{1, k}$, 
 written out in \eqref{ALmodularelliptic},
as well as the even parity property of the elliptic genus, to simplify the final result: 
\begin{align}
\chi_{L} (\tau, \a, \b)
&=  \frac{i \theta_{11}(\tau,-\alpha)}{\eta^3}
 z^{1+\frac{[k \beta_2]}{k}} {\widehat {\cal A}}_{1, k} (\tau, -\frac{\a}{k}, -2\a+k\b- k \tau -  \tau [k \beta_2])
\cr
&= \frac{i \theta_{11}(\tau,-\alpha)}{\eta^3}
{\widehat {\cal A}}_{1, k} (\tau, -\frac{\a}{k}, -2\a+k\b )
\cr
&= \frac{i \theta_{11}(\tau,\alpha)}{\eta^3}
{\widehat {\cal A}}_{1, k} (\tau, \frac{\a}{k}, 2\a -k\b ) \, .
\label{finalresult}
\end{align}
This agrees with \cite{Ashok:2011cy} and completes the identification
of the elliptic genus as the modular completed Appell-Lerch sum for
complexified chemical potentials.

We see from the final expression
that an alternative appropriate identification of variables is
\cite{Troost:2010ud, Ashok:2011cy}:
\be\label{map}
u= \frac{\a}{k} \, , \qquad v = 2\a -k\b\, .
\ee
Yet another road to the same result is obtained if one
starts out with the identification of
 variables 
\be\label{map4}
u= \frac{\a}{k} \, , \qquad v = 2\a - k \tau -k\b\, .
\ee
Note that the corresponding
Appell-Lerch sum differs from the holomorphic part
of the elliptic genus in equation (\ref{holomorphicpiece}). One then adds the difference
to the holomorphic part, and subtracts the difference from the remainder term. 
The latter term can be written as a sum over the difference of sign functions (see e.g. \cite{DMZ}).
These various forms of identifications of variables, use of the periodicity formulas, as well
as additions and subtractions code some interesting physics that we explore in the 
next subsection.

\subsection{Supersymmetric Quantum Mechanics and Wall Crossing}
We have obtained a Hamiltonian form for the elliptic genus, and
matched it onto results in mathematics. We now wish to interpret these
formulae in physical terms.  We indicate at least two interesting
phenomena. First of all we note that the final formula for the
Hamiltonian form of the elliptic genus, equations
(\ref{holomorphicpiece}) and (\ref{remainderpiece}), factorizes into a
part which we can associate to a free oscillator sum over all
generators of the $N=2$ superconformal algebra on the left, multiplied
by an Appell-Lerch sum, which is associated to weighted traces in
the radial supersymmetric quantum mechanics problems associated
to the right-moving primaries \cite{Troost:2010ud, Ashok:2011cy, Ashok:2013kk}.

Consider the remainder term (\ref{remainderpiece}) first:
\be
\chi_{L,rem}(\tau, \a, \b) = 
\frac{i\theta_{11}(\tau,-\a)}{\pi\eta^3(\tau)}\sum_{n,w}\int_{\mathbb{R}} \frac{ds}{ 2 i s + n+kw - k \beta_2}
z^{\frac{n-kw}{k}} y^{kw} q^{-nw}  (q \bar{q})^{\frac{s^2}{k}+\frac{ (n+kw-k \beta_2)^2}{4k}} \, .
\label{remaindernandw}
\ee
We note a shift
in the right-moving momentum due to the imaginary part $\beta_2$ of
the chemical potential for angular momentum. Indeed, there is an extra
term in the Lagrangian due to the angular momentum operator insertion which is
proportional (as far as the right-movers are concerned) to the
right-moving angular momentum. This term shifts the definition of the
right-moving momentum, which in turn shifts the definition of the
right-moving supercharge.  It therefore influences the measure to be
the one indicated in equation (\ref{remaindernandw}). 
  The
right-moving Hamiltonian, a function of the right-moving momentum
squared, is shifted as well, as can be seen from the exponent of
$\bar{q}$ in equation (\ref{remaindernandw}): 
\be \tilde{L}_0 -
\frac{c}{24} = \frac{s^2}{k} + \frac{(n+kw -k \beta_2)^2}{4k} \, .
\ee 
Due to the correlated shift in the right-moving supercharge and
right-moving Hamiltonian, we see that the pole contributions are still
holomorphic, right-moving ground state contributions.  The 
contributions to the left-moving momentum are fixed by the fact that a
continuous deformation keeps the locality condition
$L_0-\tilde{L}_0=-nw$ intact.\footnote{For future purposes, we have recalled
some properties of the right-moving
supersymmetric quantum mechanics in appendix \ref{SQM} (which in turn is based on 
e.g. \cite{Akhoury:1984pt}). Our discussion in words can be followed in technical
detail by making the identifications
$\tilde{L_0}-c/24=H$, $s^2/k=p^2$, $\Phi_0^2=(v-k\beta_2)^2/(4k)$
 between the variables here
and those in appendix \ref{SQM}.
As in \cite{Ashok:2013kk}, each value of the right-moving momentum $v$ gives rise
to one radial supersymmetric quantum mechanical system.}

\label{wallcrossing}

The second phenomenon, and in fact most phenomena discussed in this
paper, is a consequence of the first. We note that the imaginary part
$\beta_2$ in the angular momentum chemical potential also influences
the holomorphic contribution to the elliptic genus:
\be
\chi_{L,hol}(\tau, \a, \b)= -\sum_{m,w} \sum_{v=-(k-1)+[k \beta_2]}^{[k \beta_2]}(-1)^m q^{(m-1/2)^2/2} z ^{m-1/2}
S_{m+kw-1}
z^{v/k-2w} y^{kw} q^{kw^2-vw} \, .
\label{pickingpolesagain}
\ee
The $k$ poles in the radial momentum $s$ plane that we pick up are a function of the size
of $\beta_2$. Whenever $k \beta_2$ crosses an integer value, we will
see a subtraction and addition to our holomorphic right-moving ground
state sum, as can be seen from equation (\ref{pickingpolesagain}).

One can also understand this phenomenon from the perspective of the radial right-moving
supersymmetric quantum mechanics (see \cite{Akhoury:1984pt} and
appendix \ref{SQM}). This is a consequence of the fact that the
constant term in the supercharge, associated to the right-moving
momentum, flips sign as a function of $\beta_2$, rendering a given
ground state either normalizable or not.\footnote{ $(\partial_x +
  \Phi_0) \psi(x) = 0$ for $x \in [0,\infty]$ gives rise to a state
  state $\psi(x) \propto e^{- \Phi_0 x }$ which is normalizable or not
  depending on the sign of the constant $\Phi_0$. The conjugate supercharge
exhibits a conjugate phenomenon.}
For integer $k \beta_2$, this influences the sum over the right-moving
momentum $n+kw$, both in its upper and lower bound. The integer number
$k$ of extended $N=2$ superconformal character contributions to the
elliptic genus (and in particular the Witten index of the model)
remains unchanged under these deformations. As one radial quantum
mechanics problem looses a state, another one gains a state. Still, we
see jumps in the bound state spectrum, and in particular the R-charges
of the Ramond sector ground states whose extended characters
contribute to the holomorphic part of the elliptic genus. 

As explained below equation \eqref{holomorphicpiece}, one can in fact write the holomorphic piece schematically as follows:
\be
\chi_{L,hol} = \sum_{
\stackrel{\text{spectral}}{\text{ flow}}
}
\ \sum_{\ell=0}^{k-1}\left[\frac{i\theta_{11}(\tau, -\a)}{\eta^3(\tau)}z^{\frac{[k\b_2]-\ell}{k}}\frac{1}{1-z^{-1}} \right]_{\stackrel{\text{spectral}}{\text{ flow}}}\,.
\ee
The term in the parenthesis is an ${\cal N}=2$ superconformal character in the Ramond sector, where the R-charge of the ground state, on which the character is built, is given by $Q_R = \frac{1}{2}-\frac{\ell}{k}+\frac{[k\b_2]}{k}$. The discrete sum is over the Ramond ground states while the final sum is over all states obtained by spectral flow of the ground states by $kw$ units, with $w$ being an integer. 
This expression shows clearly that the bound state spectrum jumps across the walls where $k \beta_2$ is integer.

\section{A model for higher order Appell-Lerch sums}
\label{higherorder}
In this section, we perform a modular covariant differentiation of the
elliptic genus to obtain more general Appell-Lerch sums and their
modular completions. The
analysis carried out in the previous section will then enable us to
provide a Hamiltonian interpretation for the modular
completions of these higher order Appell-Lerch sums.

\subsection{Modular covariant derivatives}
\label{modcovder}

We confirmed in section \ref{pathintegral} that the Liouville elliptic genus
$\chi_L$ with complexified chemical
potentials is proportional to a (completed) Appell-Lerch sum.  It thus transforms as a Jacobi form under modular transformations.
In this section, we temporarily strip away the $i\theta_{11}/\eta$ prefactor
from the elliptic genus \eqref{finalresult}.
We can then study modular covariant derivatives of Appell-Lerch sums, following \cite{DMZ}.
For the isolated Appell-Lerch sum relevant to us, we have the modular transformation
property (see equation \eqref{ALmodularelliptic}):
\be 
\widehat{\cal A}_{1, k}\left(\frac{a\tau+b}{c\tau+d}, \frac{u}{c\tau+d},
\frac{v}{c\tau +d}\right) = (c\tau+d)\, e^{\frac{2\pi i c}{c\tau
    +d}(vu-ku^2) }{\widehat{{\cal A}}_{1, k}}(\tau, u, v)\, . \ee
The chemical potentials in the conformal field theory are related to the variables
$u$ and $v$ above by the relations
\be 
u=\alpha/k \, , \quad v = 2 \a - k \b \, .
\ee 
For our purposes, it is convenient to calculate in terms of
the conformal field theory variables $(\alpha,\beta)$. To make this
less cumbersome, we introduce the notation:
\begin{eqnarray}
\widehat{{\cal I}}_{1,k}  (\tau,\alpha,\beta) &=& {\widehat{{\cal A}}_{1, k}}(\tau, \frac{\alpha}{k}, 2 \alpha - k \beta)
\, .
\end{eqnarray}
The notation $\widehat{{\cal I}}$ is a reminder of the fact
 that this quantity codes generalized Witten indices of an infinite
set of radial supersymmetric quantum mechanics problems \cite{Troost:2010ud, Ashok:2011cy, Ashok:2013kk}.
In this notation, we have: 
\be\label{Imod}
 {\widehat{{\cal I}}_{1, k}}\left(\frac{a\tau+b}{c\tau+d}, \frac{\a}{c\tau+d},
\frac{\b}{c\tau +d}\right) = (c\tau+d)\, e^{\frac{2\pi i c}{c\tau
    +d}(\frac{\a^2}{k}-\a\b) }{\widehat{{\cal I}}_{1, k}}(\tau, \alpha,\beta)\, . 
\ee
This can equally well be derived from the modular transformation of the elliptic genus by
stripping away the modular properties of the theta- and eta-functions. 

To obtain a new interesting index \cite{DMZ}, we act with the 
derivative operator 
\be\label{derivative}
{\cal D} = \frac{1}{2\pi i} \left[ 2\frac{d}{d\b} + k\frac{d}{d\a} \right]
\, 
\ee
on both sides of equation \eqref{Imod} to obtain
\begin{multline}
{\cal D}\cdot\widehat{{\cal I}}_{1,k}\left(\frac{a\tau+b}{c\tau+d}, \frac{\a}{c\tau+d}, \frac{\b}{c\tau +d}\right)  
= 
(c\tau+d)e^{\frac{2\pi i  c}{c\tau +d}(\frac{\a^2}{k}-\a\b) } {\cal D}\cdot \widehat{{\cal I}}_{1,k}(\tau, \a, \b) \cr
-c\, k\b\, e^{\frac{2\pi i  c}{c\tau +d}(\frac{\a^2}{k}-\a\b) }{\widehat {\cal I}}_{1,k}(\tau, \a, \b) \, . 
\end{multline}
In order to compensate for the anomalous second term, 
consider the modular transformation property of the non-holomorphic expression -- recall that
$\beta_2 = \frac{\Im(\beta)}{\tau_2}$ --:
\begin{align}
\beta_2 & \rightarrow 
(c\tau+d)\b_2 - c \b \, .
\end{align}
It follows that the combination
\be\label{gisdf}
\widehat{{\cal I}}_{2, k}(\tau, \a,\b) =
\left( {\cal D}- k\b_2\right)\cdot \widehat{{\cal I}}_{1, k}(\tau, \a, \b)  
\ee
is a three variable function that transforms modularly:
\footnote{Other covariant differentiations to higher weight modular forms exist.}
\be
\widehat{{\cal I}}_{2,k}\left(\frac{a\tau+b}{c\tau+d}, \frac{\a}{c\tau+d}, \frac{\b}{c\tau+d}\right)= (c\tau +d)^2e^{\frac{2\pi i  c}{c\tau +d}\frac{\a^2}{k} -\a\b}{\widehat {\cal I}}_{2,k}(\tau, \a, \b)\,.
\ee
This technique was used in \cite{DMZ} to obtain higher weight
Appell-Lerch sums and their modular completions. The differentiation
process will augment the order of the denominator in the Appell-Lerch
sum.  Since the covariantization does not depend upon the weight of
the form on which the derivative acts, one can continue this process
and obtain higher weight Jacobi forms iteratively 
\be\label{gisdfforn}
\widehat{{\cal I}}_{n, k}(\tau, \a,\b) =
({\cal D}- k\b_2)\cdot \widehat{{\cal I}}_{n-1, k}(\tau, \a, \b)  \, ,
\ee
where the label $n$ specifies the weight of the modular form.

Here, we are
interested in providing a Hamiltonian interpretation for
the Jacobi forms ${\widehat {\cal I}}_{n,k}(\tau, \a)$, obtained by
setting $\b=0$, for both their
holomorphic and remainder terms. We will also provide a microscopic model
for the degrees of freedom coded in these generalized indices.

\subsection{Completed Appell-Lerch sums as state space sums}

In this section we will exhibit the decomposition of the completed
Jacobi form ${\widehat {\cal I}}_{n,k}$ into a holomorphic piece and
a remainder. We use the integral representation naturally provided by
the Liouville elliptic genus, stripped of theta- and eta-function factors.
Since the integrals allow for a radial supersymmetric quantum mechanics
interpretation, this will lead to a natural Hamiltonian interpretation for
the Appell-Lerch sums and their modular completions.

\subsubsection*{The Hamiltonian viewpoint}
Consider the Appell-Lerch sum ${\cal I}_{1,k}$ written as in equation
\eqref{holomorphicpiece}: 
\be
{\cal I}_{1, k}= 
- z^{[k \beta_2 ]/k} 
\sum_{w\in \mathbb{Z}} \frac{q^{kw^2} (z^{-2} y^{k} q^{-[ k \beta_2]})^w}{1-z^{-\frac{1}{k}}q^w}
\label{I1}\,.
\ee
The remainder term, expressed as an integration over states, is (see
equation \eqref{remainderpiece}):
\be\label{R1k}
{\cal R}_{1,k} = -\frac{1}{\pi}
\,\sum_{v\in \mathbb{Z}}\sum_{w\in \mathbb{Z}}
q^{kw^2-vw}
z^{-2w+\frac{v}{k}}
y^{kw} 
\int_{\mathbb{R} }\frac{ds}{2is +v-k \beta_2} 
(q\bar{q})^{\frac{s^2}{k} +\frac{(v-k \beta_2)^2}{4k}}
\, . 
\ee
To simplify the discussion, we choose $\beta_2$ to be in the
interval $] 0, 1/k [\,$. This is equivalent to setting $[k \beta_2]=0$
in what follows. 
This may look dangerous, since we are planning to take a derivative with respect to $\beta\,$. Note however that the sum of the terms in equations
\eqref{I1} and \eqref{R1k} behave well as functions of $\beta$, so
that we can ignore this subtlety. 
This property of continuity
  and differentiability in $\beta$ is also clear from the expression
  for the Liouville elliptic genus in path integral form.
\be 
{\cal I}_{2,k}(\tau, \a, \b) = {\cal D}\, {\cal I}_{1,k} =
\sum_{w}\frac{q^{w (k w+1)} y^{k w} z^{-\frac{1}{k}-2
    w}}{(1-z^{-\frac{1}{k}}q^w)^2} \, .  
\label{I2k}
\ee
We will refer to expression \eqref{I2k} as the double pole Appell-Lerch sum. 

Let us now act with the derivative operator on the integral representation of the
remainder:
\begin{multline}
{\cal D}{\cal R}_{1,k}
=
- \frac{1}{\pi}\,\sum_{v\in \mathbb{Z}}\sum_{w\in \mathbb{Z}}
q^{kw^2-vw}
z^{-2w+\frac{v}{k}}
y^{kw}\cr
\times \int_{\mathbb{R} } ds
\left(
  -\frac{k}{2\pi\tau_2}\frac{1}{(2is +v-k \beta_2)^2} 
+\frac{k \beta_2 }{2is +v-k \beta_2} 
\right)
(q\bar{q})^{\frac{s^2}{k} +\frac{(v-k \beta_2)^2}{4k}}   \, .
\end{multline}
The first term in the parenthesis can be rewritten using integration by parts to combine it
 with the second term: 
\be\label{DR1k}
{\cal D}{\cal R}_{1,k}
=
- \frac{1}{\pi}\,\sum_{v\in \mathbb{Z}}\sum_{w\in \mathbb{Z}}
q^{kw^2-vw}
z^{-2w+\frac{v}{k}}
y^{kw}
\int_{\mathbb{R} } ds
\left(
  -1
+\frac{v}{2is +v-k \beta_2} 
\right)
(q\bar{q})^{\frac{s^2}{k} +\frac{(v-k \beta_2)^2}{4k}} \, .
\ee
The measure in the second term is identical to that of the remainder
in the single pole case and the integral over radial momentum $s$ can be
done as before by using equation \eqref{signminuserf}.
 We will now make a brief digression to make contact with the results of \cite{DMZ}.

\subsubsection*{Relation to earlier work}
 
Setting $\b=0$ in equation \eqref{DR1k}, we obtain the remainder for ${\cal I}_{2,k}$:
\begin{eqnarray}
{\cal R}_{2,k}(\tau, \a) &=& \frac{1}{2\pi}\sqrt{\frac{k}{\tau_2}}
\sum_{v,w}q^{kw^2-vw}z^{-2w+\frac{v}{k}}
(q\bar{q})^{\frac{v^2}{4k}} \nonumber \\ 
& & -\frac{1}{\pi}\sum_{v,w}\ v\ q^{kw^2-vw}z^{-2w+\frac{v}{k}}
\, \int_{\mathbb{R}}\frac{ds}{2is+v}(q\bar{q})^{\frac{s^2}{k}+\frac{v^2}{4k}} 
\, .
\end{eqnarray}
Using the definition of the theta function with finite index and the integral
 \eqref{signminuserf}, we find
\begin{multline}
{\cal R}_{2,k}(\tau,\a) = \frac{1}{2\pi}\sqrt{\frac{k}{\tau_2}}
\sum_{v\in\mathbb{Z}}\bar{q}^{\frac{v^2}{4k}}\, 
\theta_{v,k}(\tau, -\frac{\a}{k}
)\,  
-\ \frac{1}{2}\sum_{v\in\mathbb{Z}}v\, q^{-\frac{v^2}{4k}}
\theta_{v,k}(\tau, -\frac{\a}{k}
)\left(\sgn(v) - \text{Erf}\left[\sqrt{\frac{\pi\tau_2}{k}}v\right] \right)\,, 
\end{multline}
which is equal to: 
\be\label{tocompare}
{\cal R}_{2,k}(\tau,\a) = \frac{1}{2\pi}\sqrt{\frac{k}{\tau_2}}
\sum_{v\in\mathbb{Z}}\bar{q}^{\frac{v^2}{4k}}\, \theta_{v,k}(\tau, -\frac{\a}{k}
)-\frac{1}{2}\sum_{v\in \mathbb{Z}}|v|\, q^{-\frac{v^2}{4k}}\ 
\theta_{v,k}(\tau, -\frac{\a}{k}
)\ \text{Erfc}\left[\sqrt{\frac{\pi\tau_2}{k}}|v|\right]\,.
\ee
In order to compare this 
result with the one in \cite{DMZ}, we split the variable $v\in \mathbb{Z}$ as
\be
v \equiv 2kN + \ell\,, \qquad\text{with}\quad N\in\mathbb{Z}\,,\  \ell\in\mathbb{Z}_{2k}\,,
\ee
and define
\be
\lambda\equiv\frac{v}{2k} = N + \frac{\ell}{2k}\,.
\ee
In terms of these new variables, the expression \eqref{tocompare} can be rewritten as
\begin{multline}
{\cal R}_{2,k}(\tau,\a) =k\sum_{\ell \in \mathbb{Z}_{2k}} \theta_{l,k}(\tau, -\frac{\a}{k}) \sum_{\lambda\in\mathbb{Z}+\frac{\ell}{2k}}
\left(\frac{1}{2\pi \sqrt{k \tau_2}}
\bar{q}^{k\lambda^2}
-|\lambda|\, q^{-k\lambda^2}
 \text{Erfc}\left[2|\lambda|\sqrt{\pi\tau_2 k}\right]\right)\,.
\end{multline}
This matches the remainder in \cite{DMZ} using the map between chemical potentials.

\subsubsection*{The Hamiltonian interpretation}
We return now to the Hamiltonian interpretation of the covariantly derived Appell-Lerch sums.
As recalled from \cite{Troost:2010ud, Ashok:2011cy, Ashok:2013kk} in subsection \ref{wallcrossing},
there is an interpretation of the bound state sum, as well as the remainder integral,
in terms of supersymmetric quantum mechanics systems labeled by the right-moving
momentum $v=n+kw$.\footnote{This interpretation does not depend on the free oscillator
sum $i \theta_{11}/\eta^3$ over the modes of the $N=2$ superconformal algebra. It has a
more universal character as argued in \cite{Giveon:2014hfa}. This is true even at finite
level $k$ as can be seen from the results of \cite{Ashok:2013kk}.}
It is therefore natural to express the remainder as a sum over the right-moving momentum quantum number:
\be
{\cal R}_{1,k}(\tau, \a, \b) = \sum_v S_{1,k} (\tau,\a,\b, v)\, ,
\ee
where we introduced
\be
S_{1,k}(\tau,\a,\b, v)=-\frac{1}{\pi} 
\,\sum_{w\in \mathbb{Z}}
q^{kw^2-vw}
z^{-2w+\frac{v}{k}}
y^{kw} 
\int_{\mathbb{R} }\frac{ds}{2is +v-k \beta_2} 
(q\bar{q})^{\frac{s^2}{4k} +\frac{(v-k \beta_2)^2}{4k}}
\, . 
\ee
{From} equation \eqref{DR1k}, it is clear that the covariant 
derivative acts on $S_{1,k} (v)$, which is the contribution from the continuous spectrum, as follows:
\be
({\cal D}S_{1,k})(\tau, \a, \b, v) =  v\, S_{1,k}(\tau, \a, \b, v) + Y_{2,k} (\tau, \a, \b, v) \,.
\ee
Here we have introduced the notation $Y_{2,k}$ to denote an ordinary 
partition sum obtained by integrating over the radial momentum:
\be
Y_{2,k}(\tau, \a, \beta, v) = \frac{1}{\pi}\sqrt{\frac{k}{\tau_2}}\sum_{w\in\mathbb{Z}}q^{kw^2-vw}\, z^{-2w+\frac{v}{k}}
y^{kw} (q\bar{q})^{\frac{(v-k \beta_2)^2}{4k}} \,.
\ee

We can finally write the holomorphic and remainder pieces of the completed Jacobi form ${\widehat {\cal I}}_{2,k}$ in a compact form:
\begin{align}\label{Ihat2}
{\cal I}_{2,k}(\tau, \a,\b) &=\sum_{w}\frac{q^{w (k w+1)} y^{k w} z^{-\frac{1}{k}-2 w}}{(1-z^{-\frac{1}{k}}q^w)^2}\cr
{\cal R}_{2,k}(\tau, \a, \b) &= -k\b_2\, \widehat{\cal I}_{1,k}(\tau, \a, \b) + \sum_v(v\, S_{1,k}(\tau, \a,\b, v)+Y_{2,k}(\tau, \a, \b, v)) \,.
\end{align}
Our derivation clarifies the Hamiltonian interpretation of these
expressions. The Appell-Lerch sum can be interpreted as an index sum
with a right-moving momentum insertion. This is consistent
with
the wedge sum formula for the Appell-Lerch sum \cite{DMZ}:  
\be
{\cal I}_{2,k}(\tau, \a, 0) =\left( \sum_{w\ge 0}\sum_{v\ge 0} - \sum_{w< 0}\sum_{v\le 0} \right)\, v\, q^{kw^2+v w}\, z^{-2w-\frac{v}{k}}  \,.
\ee
The first term in the remainder ${\cal R}_{2,k}$ arises from a simple
multiplication in the covariant derivative. The second term arises
from a right-moving momentum operator insertion.  The third term finds
its origin in the dependence of the spectral asymmetry as well as the 
Hamiltonian on the right-moving momentum. This fact is
explained in detail in appendix \ref{SQM}, equation
\eqref{derivativeconstant}. This indeed identifies $Y_{2,k}$ as a
partition sum of degrees of freedom living on the asymptotic cylinder.
We have thus found a Hamiltonian interpretation of the
modular double pole Appell-Lerch sum.

\subsection{Higher order Appell-Lerch sums and their completion}

We can also find a Hamiltonian integral representation 
for the remainder functions of all higher order Appell-Lerch sums. 
We act with the covariant derivative $n-1$ times to obtain the
holomorphic and remainder piece of ${\widehat {\cal I}}_{n,k}$. 
To compute the explicit expression one needs
the action of the covariant derivative on the partition sum $Y_{2,k}$, which is:
%
\be
({\cal D} - k \b_2) Y_{2,k} = 0 \, .
\ee
This knowledge is sufficient to write down the explicit integral
representations of the remainders for any order $n$ and level $k$ in terms of the integral $S_{1,k}$
over the supersymmetric quantum mechanics labeled by the right-moving momentum,
and the partition sum $Y_{2,k}$. The holomorphic parts of these higher weight
Appell-Lerch sums, when $\b=0$, are written out in \cite{DMZ} in terms
of Euler functions. Explicitly, these are given by
\begin{eqnarray}
\label{EulerAL}
\label{Inkdef}
 {\cal I}_{n+1,k}(\tau, \a)
= \left. {\cal D}^{n} {\cal I}_{1,k}(\tau, \a, \b)\right|_{\b=0}  
=
\sum_{w\in \mathbb{Z}} q^{kw^2} z^{2w}{\cal E}_{n+1}(q^w z^{-\frac{1}{k}})\,,
\end{eqnarray}
where the ${\cal E}_n$ are Euler functions, that have a series expansion of the form
\begin{align}
{\cal E}_{n+1}(x) &= 
\sum_{m>0} m^{n}x^{m} \qquad \text{if} \qquad |x| >1 \cr
&=
- \sum_{m<0} m^{n}x^{-m} \qquad \text{if} \qquad |x| <1 \,.
\end{align}
The remainders for these generalized Appell-Lerch sums become:
\be
{\cal R}_{n+1,k}(\tau, \a) =({\cal D} - k\b_2)^{n}\, \left.\widehat{\cal I}_{1,k}(\tau, \a, \b)\right\vert_{\b=0} - {\cal I}_{n+1,k}(\tau, \a) \, .
\ee
Explicitly, these are given by
\be
{\cal R}_{n+1,k}(\tau, \a) = {\cal D}^{n} {\cal R}_{1,k}\left.\right|_{\b=0} 
+\sum_{m} \frac{r_{m}}{m!}\left(\frac{k}{2\pi \tau_2}\right)^m\, {\cal D}^{n-2m}\, \widehat{\cal I}_{1,k}\left. \right|_{\b=0}
\label{rnk}
\ee
The coefficient $r_m$ is a product of binomial coefficients, given by
\be
r_m = \prod_{\ell=1}^{m}\begin{pmatrix}
n-2\ell\cr
2
\end{pmatrix}
\label{combinatorics}
\ee
The first term in equation \eqref{rnk}
corresponds to insertion of powers of the right-moving momentum,
as well as the dependence of the remainder term on the chemical potential through
the partition sum $Y_{2,k}$ and its derivatives. The list of other terms arises
from the explicit $\beta_2$ dependence of the covariant derivative, which needs to
be derived and taken into account at every given order, giving rise to the combinatorics
exhibited in equation \eqref{combinatorics}.

\subsection{Generalized elliptic genera}
In subsection \ref{modcovder}, we isolated the Appell-Lerch sum
$\widehat{I}_{1,k}$ from the elliptic genus $\chi_L$ in order to widen
the applicability of our Hamiltonian interpretation of
the holomorphic and remainder contributions to 
the higher order Appell-Lerch sums. In this subsection, we
return to the context of two-dimensional conformal field theory, in
which the radial supersymmetric quantum mechanics models have a direct
interpretation in terms of the dynamics of the right-moving
superconformal primaries. We thus wish to dress the covariant
differentiation of higher order Appell-Lerch sums with the factors
corresponding to a free $N=2$ superconformal algebra generator sum for
the left-movers.

Recall that the elliptic genus transforms as follows under a modular transformation:
\be
\chi\left(\frac{a\tau+b}{c\tau+d}, \frac{\a}{c\tau+d},\frac{\b}{c\tau+d}\right) = e^{(1+\frac{2}{k})\frac{\pi i \a^2}{c\tau+d}-\frac{2\pi i \a\b}{c\tau+d}} \chi(\tau, \a, \b)\, .
\ee
To account for the modular properties of the prefactors, we introduce a modification of the 
modular covariant derivative \eqref{derivative}:
\be
\chi^{(1)}(\tau, \a, \b) = \big({\cal D}+\a_2-k\b_2 \big) \chi(\tau, \a, \b)\, .
\ee
We then obtain a weight one Jacobi form that transforms with the same
index as the original elliptic genus. The extra factor of $\a_2$ takes
care of the anomalous transformation of the derivative of the theta
function. The technical points and Hamiltonian interpretations we
proposed in the previous subsections go through for the modularly
derived elliptic genus, with one modification. Since we have multiplied
in the theta-function which depends on the chemical potential $\alpha$,
we find terms that are associated to fermion number operator insertions. This can also be seen by recalling the definition of the elliptic genus in \eqref{EGdef}:
\be
\chi(\tau, \a, \b) = \text{Tr}_{\cal H}(-1)^{F_L +F_R} q^{L_0-\frac{c}{24}}z^{J^R_0}y^{P}\,.
\ee
A naive
action
with the differential operator ${\cal D}$ in \eqref{derivative}, yields
 the insertion
%
\be
{\cal D}\cdot \chi(\tau, \a, \b) \approx \text{Tr}_{\cal H}\left[\left(k\, J^R_0+2\, P \right)(-1)^{F_L +F_R} q^{L_0-\frac{c}{24}}z^{J^R_0}y^{P}\right]\,.
\ee
Now, the R-charge and the total angular momentum in the coset
conformal field theory can be written explicitly in terms of the
asymptotic left and right moving momenta as well as the fermion number
as follows (see for instance \cite{Israel:2004jt}):
%
\be
J^R_0 = -\frac{2}{k} P_L + F_L \qquad P = P_L + P_R \,.
\ee
Substituting this into the derivative expression, we find
\be
{\cal D}\cdot \chi(\tau, \a, \b) \approx \text{Tr}_{\cal H}\left[\left(2\, P_R + k\, F_L\right)(-1)^{F_L +F_R} q^{L_0-\frac{c}{24}}z^{J^R_0}y^{P}\right]\,.
\ee
We recognize the insertion of the right moving momentum operator we
observed in \eqref{Ihat2}; in addition we find contributions from
the fermion number insertion.
The derivation we gave here is naive, since it does not take into account
the dependence of the measure on the variables with respect to which
we derive. However, since there is no dependence on
$\alpha$ in the measure, nor in the right-moving supersymmetric
quantum mechanics, the fermion number insertion
 is the only modification in the picture we
painted previously. This is related to the holomorphic dependence of the
elliptic genus on the chemical potential for R-charge.
 We stress that in this conformal field theory
context, the origin of the radial supersymmetric quantum mechanics
systems for the right-movers can be derived from first principles.

\section{Conclusions}
\label{conclusions}

We have obtained path integral expressions for the Liouville and cigar elliptic genera with
complexified chemical potentials, and checked the modular and periodicity properties directly
from these expressions. 
We have also shown that the spectral density asymmetry depends on the imaginary part of the chemical potential $\beta$ for the angular momentum on the asymptotic circle.
We were able to exhibit a wall-crossing phenomenon in which bound
states appear and disappear from the spectrum as a function of the
imaginary part of the chemical potential.  {From} the path integral
form of the elliptic genus it is clear that the discontinuity in the
holomorphic part is mirrored in a discontinuity of the continuum
contribution such that the full expression is continuous in
$\beta$.\footnote{A similar phenomenon arises in the context of
  massive ${\cal N}=2$ theories in two dimensions
  \cite{Cecotti:1992qh} and ${\cal N}=2$ gauge theories in four
  dimensions \cite{Boristalk}.}

Furthermore, the generalization to complexified chemical potentials allowed us to take modular covariant
derivatives to obtain higher order Appell-Lerch sums and their modular completions. 
This enabled us to provide a microscopic model for higher order
Appell-Lerch sums in which we could give an interpretation to the
individual contributions as arising from bound state sums and the
continuous spectrum, whose presence results from the spectral
asymmetry. Modular derivatives moreover give rise to operator
insertions, and a new ingredient in the Hamiltonian interpretation,
which is an ordinary partition sum (with trivial spectral weight).

There are many directions for further research. One is the generalization of our analysis to the 
Jacobi forms of \cite{Ashok:2013zka,Murthy:2013mya,Ashok:2013pya}, corresponding to models with
more space-time directions (and in the gauged linear sigma-model language, to models with $N$ charged
chiral scalar fields).
The generalization of the expression \eqref{1101result} for higher $N$ is 
\begin{multline}\label{modular}
\chi_{{N}}(\tau,\a,\b_i) = \int_0^1 ds_1 ds_2 \prod_{i=1}^{N}\left[\frac{\theta_{11}(\tau,s_1\tau+s_2-\a-\frac{N\a}{k}+\b_i)}{\theta_{11}(\tau,s_1\tau+s_2-\frac{N\a}{k}+\b_i)}\right]\cr
\times \sum_{m,w \in \mathbb{Z}}e^{-2\pi i s_2 w}e^{2\pi i s_1(m-N\a)}
e^{-\frac{\pi}{k\tau_2}|m-N\a+w\tau|^2}\, .
\end{multline}
One can check that it is modular covariant and elliptic, with central charge $c=3N(1+\frac{2N}{k})\,$. We can  obtain the analogue of the path integral expression 
\eqref{chicoss1s2} via double Poisson resummation:
\begin{multline}
\chi_{{N}}(\tau,\a,\b_i) = k
\int_{0}^1 ds_1 ds_2  \prod_{i=1}^N \left[\frac{\theta_{11}(\tau,s_1\tau+s_2-\a-\frac{N\a}{k}+\beta_i)}{\theta_{11}(\tau,s_1\tau+s_2-\frac{N\a}{k}+\beta_i)}\right]\, 
\cr \times
\sum_{m,w \in \mathbb{Z}}e^{2\pi i N \a w}\,
e^{- 2 \pi i N \alpha_2( (s_1 +w) \tau + s_2 +m)} \, 
e^{-\frac{\pi k}{\tau_2}|m+w \tau +(s_1\tau+s_2 )|^2}\, .
\end{multline}
This is a good starting point for an analysis of the generalized Hamiltonian interpretation, covariant
differentiation, etc. A similar analysis can also be carried out for orbifold and tensor product models.

Another direction for future research was one of the
original motivations of this paper. We have given a microscopic
interpretation of the higher level Appell-Lerch sums, dressed with
further theta- and eta-functions.  It should be clear from our
analysis that the radial quantum mechanics giving rise to the
Appell-Lerch sums exhibits universal features independent of the
particular dressing factor. We hope that these features 
will provide useful hints towards a
microscopic interpretation of the mock Jacobi forms arising in the context
of microscopic black hole entropy counting,
and in particular the modularly completed 
single or multi-centered black hole partition sums of \cite{DMZ}.

\section*{Acknowledgments}
We would like to thank Suresh Nampuri, Boris Pioline, 
Giuseppe Policastro, Ashoke Sen and especially Atish Dabholkar and Sameer Murthy for useful discussions.

\begin{appendix}

\section{The path integral of the axially gauged coset model}\label{cosetpathintegral}

In this appendix we show how to obtain the path integral expression for the elliptic genus when the chemical potentials $\a$ and $\b$ are taken to be complex. As shown in \cite{Ashok:2011cy} the path integral can be written in the factorized form 
\be
\chi^{PI}(\tau,\a) = \int_{\mathbb{C}}\frac{d^2u}{2i\tau_2}\ Z_{g}(\tau,\a) \, Z_{f}(\tau,\a) \, Z_{Y}(\tau,\a) \, Z_{gh}(\tau)\, Z_{a}(\tau,\a) \,.
\ee
We follow the notations and conventions of \cite{Ashok:2011cy} and refer to that reference for 
details. The subscripts denote the various sectors: the
$SL(2,\mathbb{R})$ group, the fermions, a compact boson $Y$ and the
ghosts. The very last contribution is an anomalous contribution that
results from a $U(1)_R$ rotation of the fermions due to their twisted
boundary conditions. In the above equation,
$u=s_1\tau+s_2$ is the complexified holonomy that takes values over
the entire complex plane.

Here, we will use a slightly different method to obtain the twisted
partition functions\footnote{The method is close to the one used in \cite{Kraus:2006nb},
for instance.}. We first write down a modular
invariant expression for the path integral in each sector. 
These are straightforward generalizations of the
expressions in \cite{Ashok:2011cy}, with the modification that we make
the contribution from each sector 
modular invariant
\begin{align}
Z_{g}(\tau, \a) &= \frac{\sqrt{k}\kappa}{\sqrt{\tau_2}}\frac{e^{\frac{2\pi}{\tau_2}(\text{Im}u-\frac{\text{Im}\a}{k})^2} }{|\theta_{11}\tau, u-\frac{\a}{k}|^2}\cr
Z_{f}(\tau, \a)&= \frac{1}{\kappa |\eta(\tau)|^2}\left[e^{-\frac{\pi}{\tau_2}(|u-\a-\frac{\a}{k}|^2-(u-\a-\frac{\a}{k})^2)}\theta_{11}(\tau, u-\a-\frac{\a}{k})\right]
\cr
& \qquad \quad \times \quad \left[ e^{-\frac{\pi}{\tau_2}(|u-\frac{\a}{k}|^2-(\bar{u}-\frac{\bar{\a}}{k})^2)}\theta_{11}(\bar{\tau}, \bar{u}-\frac{\bar{\a}}{k})\right]\cr
Z_{Y}(\tau,\a) &= \sqrt{\frac{k}{\tau_2}}\frac{1}{|\eta(\tau)|^2}\, e^{-\frac{k\pi}{\tau_2}|u|^2}\cr
Z_{gh}&=\tau_2|\eta(\tau)|^4 \,.
\end{align}
The further anomaly factor results from a chiral rotation of the
fermions that is necessary to write the action as a sum over independent
sectors; we find
\begin{align}
Z_{a} &= e^{\frac{\pi }{2\tau_2} (u\bar{\alpha}- \bar{u}\alpha) } 
\, .
\end{align}
This is, once again, a generalization of the result in \cite{Ashok:2011cy}, now taking the chemical potentials to be complex. The exponent can be understood as coming from a chiral rotation of the fermions that depends on the chemical potential $\a$, which measures the R-charge. This leads to an anomaly exponent of the form
\be
\int A^R \wedge dY^u\, ,
\ee
where $A^R$ is the background gauge field coupling to the R-charge and $d Y^u$ is the differential
of the twisted compact boson.
Making use of the Riemann bi-linear identity and the definition of the holonomies, we obtain the quoted result.  Putting all this together, we find
\be
\chi^{PI}(\tau,\a) = e^{\frac{ \pi \hat{c}(\a^2-|\a|^2)}{\tau_2}}\, k\, \int_{\mathbb{C}}\frac{d^2u}{2i\tau_2} 
\frac{\theta_{11} (u- \frac{k+1}{k} \alpha )}{
\theta_{11}(u - \frac{\alpha}{k}  )} 
e^{ - \frac{k \pi}{ \tau_2} |u |^2} 
 e^{-2\pi i\frac{\a_2}{\tau_2}u} \, .\,  
\ee
One can check that the $\a$-dependent prefactor is such that under a modular transformation, the path integral elliptic genus is modular invariant. The Hamiltonian expression for the elliptic genus is simply obtained by omitting this prefactor (see e.g. \cite{Kraus:2006nb}); we finally find
\be
\chi(\tau,\a) =k\, \int_{\mathbb{C}}\frac{d^2u}{2i\tau_2} 
\frac{\theta_{11} (u- \frac{k+1}{k} \alpha )}{
\theta_{11}(u - \frac{\alpha}{k}  )} 
e^{ - \frac{k \pi}{ \tau_2} |u |^2} 
 e^{-2\pi i\frac{\a_2}{\tau_2}u} \, .\,  
\ee
This calculation can easily be generalized to include the chemical potential $\beta$.

\section{Towards the Hamiltonian viewpoint}
\label{lagrangiantohamiltonian}

In section \ref{pathintegral}, we obtained the path integral form of the Liouville elliptic genus
$\chi_L$ with complexified chemical
potentials $\alpha$ and $\beta$
\begin{multline}
\chi_L(\tau, \a, \b)=
\sum_{n,m} \int_0^1 ds_{1} ds_{2}
\frac{\theta_{11} (\tau,s_1\tau +s_2 - \alpha )}{
\theta_{11}(\tau, s_1 \tau+ s_2 )} e^{ 2 \pi i \alpha \frac{n}{k}}
e^{ - \frac{k \pi}{ \tau_2} | (\frac{n}{k}+ s_1 ) \tau + (\frac{m}{k}+ s_2 ) + \frac{\alpha}{k} -  \beta |^2} 
\cr 
\times e^{-2\pi i\a_2((\frac{n}{k}+ s_1 ) \tau + (\frac{m}{k}+ s_2 ) +\frac{\alpha}{k} -  \beta)}\, .
\label{startingpointappendixB}
\end{multline}
In this appendix we will show how to rewrite this in a form that lends
itself to a Hamiltonian interpretation, following \cite{Troost:2010ud, Ashok:2011cy} closely. 
We perform Poisson resummation
on the $m$ quantum number in equation \eqref{startingpointappendixB} and find:
\begin{multline}\label{zwegersexactly}
\chi_L(\tau, \a, \b) = \sqrt{k \tau_2} 
\sum_{n,w} \int_0^1 ds_{1} ds_{2}
\frac{\theta_{11} (\tau,s_1\tau +s_2 - \alpha )}{
\theta_{11}(\tau, s_1 \tau+ s_2 )} e^{ 2 \pi i \alpha \frac{n}{k}}\, e^{- 2 \pi i w (k s_2 + \alpha -k \beta_1)}
\cr
\times q^{ (kw-(n+k s_1-k \beta_2))^2/4k} \bar{q}^{(kw + (n+ k s_1-k \beta_2))^2/4k} \, .
\end{multline}
We have shifted the complexified chemical potentials into the exponents
to easily expand the denominator theta function following
 \cite{Troost:2010ud, Ashok:2011cy}. 
Using the expansions in equations \eqref{thetaexp} and \eqref{invthetaexp}, we obtain
\begin{multline}
\chi_L (\tau, \a, \b) = -\sqrt{k \tau_2} 
\frac{1}{\eta^3} \sum_{m,r,n,w} \int_0^1 ds_{1} ds_{2}
(-1)^m q^{(m-1/2)^2/2} (z e^{- 2 \pi i s_2} q^{-s_1})^{m-1/2}
( e^{2 \pi i s_2} q^{s_1} )^{r+1/2} 
\cr
S_r(q)
z^{n/k} e^{- 2 \pi i w (k s_2 + \alpha -k \beta_1)} q^{ (kw-(n+k s_1 - k \beta_2))^2/4k} \bar{q}^{(kw + (n+ k s_1- k \beta_2))^2/4k} \, .
\end{multline}
The holonomy integral over $s_2$ imposes the Gauss constraint $r-m+1=kw\,$, leading to the simplified expression
\begin{multline}
\chi_L (\tau, \a, \b) = -\sqrt{k \tau_2} 
\frac{1}{\eta^3} \sum_{m,n,w} \int_0^1 ds_{1} 
(-1)^m q^{(m-1/2)^2/2} z ^{m-1/2} S_{m+kw-1}(q)  
\cr
z^{n/k} e^{- 2 \pi i w (\alpha -k \beta_1)} q^{kw s_1} q^{ (kw-(n+k s_1- k \beta_2))^2/4k} \bar{q}^{(kw + (n+ k s_1 - k \beta_2))^2/4k} 
\, .
\end{multline}
Since we know that the power of $\bar{q}$ will be zero for the right-moving ground states, it is advantageous to recombine the exponents of the last two factors as follows:
\be
q^{ (kw-(n+k s_1- k \beta_2))^2/4k} \bar{q}^{(kw + (n+ k s_1 - k \beta_2))^2/4k} =q^{-nw+kw(\b_2-s_1)} (q\bar{q})^{(kw + n+ k s_1 - k \beta_2)^2/4k} \, .
\ee
The elliptic genus then simplifies to:
\begin{multline}
\chi_L (\tau, \a, \b)= -\sqrt{k \tau_2} 
\frac{1}{\eta^3} \sum_{m,r,n,w} \int_0^1 ds_{1} 
(-1)^m q^{(m-1/2)^2/2} z ^{m-1/2}
 S_{m+kw-1}(q)  
\cr
z^{n/k-w} y^{kw}   q^{-nw}  (q \bar{q})^{(kw + n+ k s_1- k \beta_2)^2/4k} \, .
\end{multline}
Next we define the right-moving momentum variable $v\equiv n+kw$ to obtain:
\begin{multline}
\chi_L (\tau, \a, \b) = -\sqrt{k \tau_2} 
\frac{1}{\eta^3} \sum_{m,r,v,w} \int_0^1 ds_{1} \delta_{r-m+1-kw}
(-1)^m q^{(m-1/2)^2/2} z ^{m-1/2}
 S_r (q)
\cr
z^{v/k-2w} y^{kw} q^{kw^2-vw}  (q \bar{q})^{(v+ k s_1- k \beta_2)^2/4k} \, .
\end{multline}
In order to linearize the exponent in $s_1$, we use the familiar trick of introducing a new variable $s$, which will later play the role of the momentum along the radial direction:
\begin{multline}
\chi_L (\tau, \a, \b) = - 
\frac{ 2  \tau_2}{\eta^3} \sum_{m,r,v,w} \int_0^1 ds_{1}\   \int_{- \infty}^{+\infty} ds\ 
(-1)^m q^{(m-1/2)^2/2} z ^{m-1/2}
 S_{m+kw-1}  (q)
\cr
z^{v/k-2w} y^{kw}  q^{kw^2-vw}  (q \bar{q})^{s_1 (is + \frac{v}{2} -\frac{k \beta_2}{2}) + \frac{s^2}{k}+ \frac{(v-k \beta_2)^2}{4k}} \, .
\end{multline}
This in turn permits us to calculate the integral over the holonomy $s_1$ in a convenient form:
\begin{multline}\label{chiinter}
\chi_L (\tau, \a, \b) = 
\frac{1}{\pi \eta^3} \sum_{m,v,w}  \int_{- \infty}^{+\infty} \frac{ds}{ 2 i s + v - k \beta_2}
(-1)^m q^{(m-1/2)^2/2} z ^{m-1/2}
 S_{kw+m-1}  (q)
\cr
z^{v/k-2w} y^{kw} q^{kw^2-vw}  [(q \bar{q})^{(is + \frac{v}{2} - \frac{k \beta_2}{2})}-1](q \bar{q})^{\frac{s^2}{k}+\frac{ (v-k \beta_2)^2}{4k}} \, .
\end{multline}
As in \cite{Troost:2010ud,Ashok:2013zka}, there is a simple way to
extract the holomorphic piece and remainder term from this expression. We
begin with the ``$1$" term in the square parenthesis. In this term, we do a combined shift of variables:
\be
s\longrightarrow s+\frac{ik}{2} \qquad v\longrightarrow v+k \,,
\ee
which leads to the following expression: 
\begin{multline}
\chi_L^{1} (\tau, \a, \b) = 
-\frac{1}{\pi \eta^3} \sum_{m,v,w}  \int_{\mathbb{R}-\frac{ik}{2}} \frac{ds}{ 2 i s + v - k \beta_2}
(-1)^m q^{(m-1/2)^2/2} z ^{m-1/2}
 S_{kw+m-1}  (q) zq^{-kw}	
\cr
z^{v/k-2w} y^{kw} q^{kw^2-vw}  (q \bar{q})^{(is + \frac{v}{2} - \frac{k \beta_2}{2})+\frac{s^2}{k}+\frac{ (v-k \beta_2)^2}{4k}} \, .
\end{multline}
Now, we use the relation
\begin{align}
S_{m+kw-1}(q) &= 1- S_{-m-kw}(q)\,.
\end{align}
The part proportional to ``$1$" we again split off and we refer to it as
the remainder; we will deal with this piece later. What remains of the term we
denote $\chi_L^{1,S}$ and it equals
\begin{multline}
\chi_L^{1,S} (\tau, \a, \b) = 
\frac{1}{\pi \eta^3} \sum_{m,v,w}  \int_{\mathbb{R}-\frac{ik}{2}} \frac{ds}{ 2 i s + v - k \beta_2}
(-1)^m q^{(m-1/2)^2/2} z ^{m-1/2}
 zq^{-kw}	\,   S_{-m-kw} (q)
\cr
z^{v/k-2w} y^{kw} q^{kw^2-vw}  (q \bar{q})^{(is + \frac{v}{2} - \frac{k \beta_2}{2})+\frac{s^2}{k}+\frac{ (v-k \beta_2)^2}{4k}} \, .
\end{multline}
We now define $\tilde{m} = m+1$, which allows us to absorb the extra
$z$ factor into the $\tilde{m}$ summation. It also leads to an extra
overall sign factor. Moreover there is an extra $q$-dependent factor
given by $q^{1-\tilde{m}}$; putting all this together we find (after
omitting the tilde on $m$)
\begin{multline}
\chi_L^{1,S} (\tau, \a, \b) = 
-\frac{1}{\pi \eta^3} \sum_{m,v,w}  \int_{\mathbb{R}-\frac{ik}{2}} \frac{ds}{ 2 i s + v - k \beta_2}
(-1)^m q^{(m-1/2)^2/2} z ^{m-1/2}
q^{-m +1 - kw}\, S_{-m-kw+1}
\cr
z^{v/k-2w} y^{kw} q^{kw^2-vw}  (q \bar{q})^{(is + \frac{v}{2} - \frac{k \beta_2}{2})+\frac{s^2}{k}+\frac{ (v-k \beta_2)^2}{4k}} \, .
\end{multline}
We can now use another relation that the series $S$ satisfies
\be
q^r S_r(q) = S_{-r}(q)\,.
\ee
This allows us to write
\begin{multline}
\chi_L^{1,S} (\tau, \a, \b) = 
-\frac{1}{\pi \eta^3} \sum_{m,v,w}  \int_{\mathbb{R}-\frac{ik}{2}} \frac{ds}{ 2 i s + v - k \beta_2}
(-1)^m q^{(m-1/2)^2/2} z ^{m-1/2}
 S_{m+kw-1}
\cr
z^{v/k-2w} y^{kw} q^{kw^2-vw}  (q \bar{q})^{(is + \frac{v}{2} - \frac{k \beta_2}{2})+\frac{s^2}{k}+\frac{ (v-k \beta_2)^2}{4k}} \, .
\end{multline}
What is remarkable is that \eqref{chiinter} can be combined with
$\chi^{1,S}$ above to give a contour integral, which we denote
\begin{multline}\label{appchiholo}
\chi_{L,hol}= \frac{1}{\pi\eta^3}\sum_{m,v,w}\left[\int_{\mathbb{R}}-\int_{\mathbb{R}-\frac{ik}{2}} \right]
\frac{ds}{ 2 i s + v - k \beta_2}
(-1)^m q^{(m-1/2)^2/2} z ^{m-1/2}
S_{m+kw-1}
\cr
z^{v/k-2w} y^{kw} q^{kw^2-vw}  (q \bar{q})^{(is + \frac{v}{2} - \frac{k \beta_2}{2})+\frac{s^2}{k}+\frac{ (v-k \beta_2)^2}{4k}} \, .
\end{multline}
This is a holomorphic contribution, since, at the location of the
poles, the exponent of the non-holomorphic piece vanishes.
The piece that is left over will be denoted the remainder and is the ``$1$" term in $\chi^{1}$, given by
\begin{multline}
\chi_{L, rem} = 
-
\frac{1}{\pi\eta^3}\sum_{m,v,w}\int_{\mathbb{R}-\frac{ik}{2}} \frac{ds}{ 2 i s + v - k \beta_2}
(-1)^m q^{(m-1/2)^2/2} z ^{m-1/2}
z q^{-kw}
\cr
z^{v/k-2w} y^{kw} q^{kw^2-vw}  (q \bar{q})^{(is + \frac{v}{2} - \frac{k \beta_2}{2})+\frac{s^2}{k}+\frac{ (v-k \beta_2)^2}{4k}} \, .
\end{multline}
To obtain the state sum interpretation,
we translate the $s$-contour back onto the real axis using the
inverse transformations
\be
s \longrightarrow  s-\frac{ik}{2} \qquad v\longrightarrow v-k \,.
\ee
This also has the effect of removing the imaginary piece from the
$(q\bar{q})$ exponent along with the left over $zq^{-kw}$ factor. As a
result we obtain (after suitably relabeling the variables):
\be
\chi_{L,rem} = 
-\frac{1}{\pi\eta^3}\sum_{m,v,w}\int_{\mathbb{R}} \frac{ds}{ 2 i s + v - k \beta_2}
(-1)^m q^{(m-1/2)^2/2} z ^{m-1/2}
z^{v/k-2w} y^{kw} q^{kw^2-vw}  (q \bar{q})^{\frac{s^2}{k}+\frac{ (v-k \beta_2)^2}{4k}} \, .
\ee
Using the theta expansion in \eqref{thetaexp}, this can be written as
\be
\chi_{L,rem} = 
\frac{i\theta_{11}(\tau,-\a)}{\pi\eta^3}\sum_{v,w}\int_{\mathbb{R}} \frac{ds}{ 2 i s + v - k \beta_2}
z^{v/k-2w} y^{kw} q^{kw^2-vw}  (q \bar{q})^{\frac{s^2}{k}+\frac{ (v-k \beta_2)^2}{4k}} \, .
\label{remainderintegral}
\ee
The final result for the elliptic genus is therefore a sum of two contributions, in \eqref{appchiholo} and \eqref{remainderintegral}. We relate these results to the mathematics of completions of Appell-Lerch sums
in the bulk of the paper.

\section{Supersymmetric quantum mechanics}
\label{SQM}

Suppose we have a quantum-mechanical system with Hamiltonian $H$:
\be
H = p^2 + \Phi^2 - [ \psi^\dagger , \psi ] \Phi'(x)
\ee
where $\Phi(x)$ is a function of the variable $x$ parameterizing the manifold
on which the supersymmetric particle propagates. The commutation relations:
\begin{align}
[ p,x ] &= -i\qquad
\{ \psi^\dagger , \psi \} = 1 
\end{align}
lead to a supersymmetry algebra:
\be
\{ Q, Q^\dagger \} = H
\ee
for the supercharges
\begin{align}
Q &= ( p + i \Phi) \psi^\dagger
\cr
Q^\dagger &= ( p - i \Phi) \psi
\, .
\end{align}
The Witten index in models with continuous part to their spectrum is given by:
\begin{multline}
Tr (-1)^F e^{- \beta H}
= N_{bos} (E=0) - N_{ferm}(E=0)   \cr
+ \int_{E_{treshhold}}^{+\infty}
d E e^{- \frac{ E}{T}} ( \rho_{bos}(E)-\rho_{ferm}(E)) \, .
\end{multline}
If we assume the quantum mechanics to live on a half-line (e.g. because of a 
steep potential at $x=0$ or $x=-\infty$), then we have a scattering problem
with bosonic and fermionic waves bouncing of the wall. The supercharge relates
the wave-functions of these excitations near $x=+\infty$ (where the potential
is assumed to take the constant value $\Phi_0$):
\begin{eqnarray}
\Psi_{bos} (x) & \propto & e^{i k x} + a_{bos}(k) e^{-i kx}
\nonumber \\
\Psi_{ferm} (x) & \propto &   e^{i k x} + a_{ferm}(k) e^{-i kx}
\nonumber \\
& \propto &
(ik + \Phi_0) e^{i k x} + (-ik + \Phi_0) a_{bos}(k) e^{-i kx} \, .
\end{eqnarray}
That leads to a difference in spectral densities equal to:
\begin{align}
\rho_{bos}(k) - \rho_{ferm}(k) &= \frac{1}{2 \pi i} \frac{d}{dk} \log \frac{a_{bos}}{a_{ferm}}
\cr
&= \frac{1}{2 \pi } \left(\frac{1}{ik + \Phi_0}- \frac{1}{ik-\Phi_0}\right ) \, .
\end{align}
{From} the behaviour of the potential at infinity, we find that the continuum of states
starts at an energy $E=\Phi_0$. We can write the (generalized) Witten index as:
\begin{align}
Tr (-1)^F e^{- \frac{ H}{T}}
&= N_{bos} (E=0) - N_{ferm}(E=0) \ + \int_{0}^{+\infty}
\frac{d k}{2 \pi }  \left(\frac{1}{ik + \Phi_0}- \frac{1}{ik-\Phi_0}\right )   e^{- \frac{ E(k^2)}{T}} 
\cr
&= N_{bos} (E=0) - N_{ferm}(E=0) + \int_{- \infty}^{+\infty}
\frac{dk}{2 \pi }  \frac{1}{ik + \Phi_0}  e^{- \frac{ E(k^2)}{T}}  
 \, .
\end{align}
The conserved energy can be evaluated at infinity to be $E=p^2 + \Phi_0^2$,
such that this simplifies further to:
\be
Tr (-1)^F e^{- \frac{ H}{T}}
= N_{bos} (E=0) - N_{ferm}(E=0)   + \int_{- \infty}^{+\infty}
\frac{dk}{2 \pi }  \frac{1}{ik + \Phi_0}  e^{- \frac{1}{T} (k^2+\Phi_0^2)}  
 \, .
\ee
Up to here, the analysis is a minor variation of \cite{Akhoury:1984pt}.
What we wish to stress is that, when we derive this weighted trace with 
respect to the constant $\Phi_0$, we find:
\be
\frac{d}{d \Phi_0} Tr (-1)^F e^{- \frac{ H}{T}}
= \int_{- \infty}^{+\infty}
\frac{dk}{2 \pi }  \left(-\frac{1}{(ik + \Phi_0)^2}- \frac{2}{T} \frac{ \Phi_0}{ik + \Phi_0}\right)  e^{- \frac{1}{T} (k^2+\Phi_0^2)}  
 \, ,
\ee
which through partial integration becomes:
\begin{align}
\frac{d}{d \Phi_0}\left[ Tr (-1)^F e^{- \frac{H}{T}}\right]
&=  - \frac{1}{\pi T } \int_{- \infty}^{+\infty}
dk  e^{- \frac{1}{T} (k^2+\Phi_0^2)}  \cr
&=- \frac{1}{\sqrt{\pi T}}\ e^{- \frac{\Phi_0^2}{T} }
 \, . 
\label{derivativeconstant}
\end{align}
Thus, the derivative with respect to the constant in the potential is
proportional to an integral over the bosonic continuum, weighted as an
ordinary partition function.

\section{Useful formulae}\label{usefulformulae}

The theta function has a power series expansion
\be\label{thetaexp}
\theta_{11}(\tau, \a) =-i \sum_{m\in \mathbb{Z}}(-1)^{m} q^{\frac{(m-\frac{1}{2})^2}{2}}z^{m-\frac{1}{2}}
\, .
\ee
The modular and elliptic properties of the theta function are given as follows:
\begin{align}\label{thetamodell}
\theta_{11}(-\frac{1}{\tau},\frac{\a}{\tau}) &= -i (-i\tau)^{\frac{1}{2}}e^{\frac{\pi i \a^2}{\tau}}\, \theta_{11}(\tau, \a) \cr
\theta_{11}(\tau, \a+ m\tau + n) &= (-1)^{m+n}\, q^{-\frac{m^2}{2}}\, z^{-m}\, \theta_{11}(\tau, \a) \,.
\end{align}
For the variable $s_1$ in the interval $s_1 \in {[} 0, 1 {]}\,$, the inverse theta function can be expanded as
\begin{eqnarray}\label{invthetaexp}
\frac{\eta^3(\tau)}{i\theta_{11}(\tau, s_1\tau+s_2)} &=& \sum_{r\in \mathbb{Z}} ( e^{2 \pi i s_2} q^{s_1} )^{r+1/2} S_r(q)\ , 
\nonumber \\
\text{with}\qquad S_r(q) &=& \sum_{n=0}^{\infty}(-1)^n\, q^{\frac{n(n+2r+1)}{2}} \,.
\end{eqnarray}
Another useful identity is
\be\label{thetabypole}
\frac{i\theta_{11}(\tau, \a)}{1-zq^{p}} = \sum_{m\in \mathbb{Z}}(-1)^mq^{\frac{1}{2}(m-\frac{1}{2})^2}z^{m-\frac{1}{2}}S_{-m+p}(q)\, .
\ee
See e.g. \cite{Ashok:2013zka} for proofs of these relations. 
We also use the theta series with finite index, defined as follows:
\be\label{thetark2}
\sum_{j\in \mathbb{Z}} q^{k(j+\frac{r}{2k})^2} 
z^{2k(j+\frac{r}{2k})} = \theta_{r,k}(\tau, \a ) \, .
\ee
%
We have made use of the integral:
\be\label{signminuserf}
\sgn(r+\gamma_2) - \text{Erf}\left(\sqrt{\frac{\pi\tau_2}{k}}(r+\gamma_2)\right) = \frac{2}{\pi} \int\frac{ds}{2is+r+\gamma_2}e^{-\frac{\pi\tau_2}{k}(4s^2+(r+\gamma_2)^2)} \, .
\ee

\end{appendix}

\end{document}